\documentclass[pra,aps,twocolumn,showpacs]{revtex4}

\usepackage[english]{babel}
\usepackage{amsmath}                             
\usepackage{graphicx}                            
\usepackage{bm}                                  
\usepackage{bbm}                                 

\DeclareBoldMathCommand{\bfrho}{\rho}
\newcommand{\ket}[1]{{#1}}                     
\newcommand{\Mpx}{M} 
\newcommand{\LBdG}{\mathcal{L}} 
\newcommand{\HGP}{\mathcal{H}} 

\begin{document}


\title{Quantum fluctuations in the image of a Bose gas}


\author{Antonio~Negretti$^1$}\email[E-mail: ]{negretti@phys.au.dk}
\author{Carsten Henkel$^2$}\author{Klaus M\o lmer$^1$}
\affiliation{1. Lundbeck Foundation Theoretical Center for Quantum System
Research\\
Department of Physics and Astronomy, University of Aarhus,
8000 Aarhus C, Denmark\\
2. Institut f\"{u}r Physik und Astronomie, Universit\"at Potsdam, 
Karl-Liebknecht-Str. 24-25, 14476 Potsdam, Germany}


\begin{abstract}
We analyze the information content of density profiles 
for an ultracold Bose gas of atoms and extract resolution limits 
for observables contained in these images. 
Our starting point is density correlations 
that we compute within the
Bogoliubov approximation, taking into account quantum and thermal 
fluctuations beyond mean-field theory.
This provides an approximate way to construct the joint counting
statistics of atoms in an array of pixels covering the gas. 
We derive the Fisher
information of an image and the associated Cram\'er-Rao sensitivity
bound for measuring observables
contained in the image.
We elaborate on our recent study on position 
measurements of a dark soliton [Negretti \emph{et al.}, Phys.\ Rev.\ A \textbf{77}, 043606 (2008)]
where a sensitivity scaling with the atomic density as $n^{-3/4}$ was 
found. We discuss here a wider class of soliton
solutions and present a detailed analysis of the Bogoliubov
excitations and the gapless (Goldstone) excitation modes.  
These fluctuations around the mean field contribute to the
noise in the image, and we show how they can actually improve the
ability to locate the position of the soliton.
\end{abstract}


\maketitle


\section{Introduction}

de Broglie waves of massive particles
are very sensitive to perturbations and
may serve as efficient probes for electromagnetic fields
\cite{Vengalattore2007}, earth's rotation \cite{Stringari2001},
Casimir forces \cite{Obrecht2007} (or in general to detect weak 
forces \cite{Bhongale2008}), and particle properties as, for
example, the refraction index of a buffer gas \cite{Schmiedmayer1995} or 
the electric polarizability of an atom \cite{Ekstrom1995}.  In interferometry,
disturbance of the phase of light or matter waves in one arm of the
interferometer can be measured by a displacement of the interference
fringes with a sensitivity determined by the fringe wavelength and the
signal-to-noise ratio (SNR).  The SNR is under many
circumstances given by the standard shot noise, leading to a
resolution that scales with $1/\sqrt{N}$, where $N$ is the number of
detected atoms. 

The shot noise limit, however, can be beaten with entanglement and 
squeezing \cite{Giovannetti2004}, proving that it is not a 
fundamental limit and in the scenario of Ref.\cite{Pezze2007}, for example, 
a particular entangled preparation scheme has been shown to give
a phase error scaling as $1/N^{3/4}$, while in principle the Heisenberg 
limit should provide the optimum sensitivity with a phase error scaling 
as $1/N$ \cite{Bouyer1997}.
Many analyses of nonclassical metrology with quantum objects have
dealt with the situation of particles or fields that may be prepared
in suitably entangled states, and entangled measurements may be used
after they have experienced the interaction of interest~\cite{Giovannetti2006}.
Conversely, entanglement created by many-body interactions improves 
the precision of estimating the corresponding coupling constant,
as discussed recently \cite{Boixo2008a,Boixo2008b}.

In this paper we arrive
at a $1/N^{3/4}$  scaling for a displacement measurement with
particles that are, however, not
entangled at all.  We consider a system of bosonic atoms,
which are cooled to Bose-Einstein degeneracy
and prepared in a joint collective quantum state,
described by mean field theory, i.e., by a Hartree-Fock
product state wave function.  
We take for the latter a dark soliton,
a topological excitation stabilized by 
atom-atom interactions.
The soliton has a density minimum at a
location $q$, and this position can be used to monitor the phase acquired 
in an atomic matter wave interferometer in the nonlinear regime 
\cite{Scott1998,Negretti2004,Scott2008}.
A recent experiment has demonstrated the relevance of the relative 
phase in splitting and recombining a Bose-Einstein condensate (BEC), 
although the direct observation of solitons was not 
possible \cite{Jo2007}.
In a similar way, very recent experiments have 
generated solitons that
oscillate and collide in a harmonic 
trap, in the crossover regime between one dimension (1D) and three dimensions 
(3D) \cite{Weller2008}. 
As we have shown in a recent 
publication \cite{Negretti2008a}, the decrease of the soliton width with 
the number of atoms (a nonlinear effect)
gives rise to an improved position
resolution, surpassing
the shot noise threshold, when more and more particles are used for the 
experiment. 

Atom clouds are typically analyzed by taking (pixelized)
absorption or phase contrast images, giving access to the atomic
density profile. 
If one wants to measure some 
quantity of interest, the resolution is limited by
the fluctuations in the image due to
counting noise on every pixel. We characterize these 
fluctuations and specify how to extract, in an (almost) optimal way,
information from the noisy data. It turns out that the quantum
fluctuations beyond the Hartree-Fock
product state actually do not spoil the image resolution, but even
improve it slightly, although the power law scaling with the atomic 
density ($\sim n^{-3/4}$) is unchanged.
In this paper, we present a description of these elements of the theory, 
providing a number of technical ingredients and details that were only briefly 
mentioned in Ref.\cite{Negretti2008a}. 

In Sec.\ref{s:information}, we introduce the  quantum image of an atomic cloud, 
define the associated Fisher information (FI) and recall its
connection to optimal parameter estimation, starting from the image data.
We treat solitons and vortices in a mean field description with Poissonian counting 
statistics, and we introduce a general Gaussian Ansatz for the 
counting statistics 
within an image, for which the Fisher information can be determined 
analytically.  
Different scalings with the atomic density are 
identified that range between the shot-noise and the Heisenberg limits, depending 
on the system dimension and the kind of nonlinearity.
In Sec.\ref{s:qft}, a quantum field theory of atomic density correlations
is developed within the Bogoliubov approximation. We provide a discussion 
of the role of phonon and zero (or Goldstone) modes, and we present detailed 
calculations for the density fluctuations in the image of a dark soliton. 
The location of the soliton is, due to the breaking of translational symmetry 
of the problem, itself associated with a Goldstone mode, and the corresponding 
contribution to particle fluctuations are analyzed. In this section, 
we also construct a nearly optimal protocol for image processing. 
In Sec.\ref{s:conclusion} we 
give a brief summary and conclusion.

\section{Mean field images}
\label{s:information}

\subsection{Atom density statistics}
\label{s:basics}

The continuous popularity of cold quantum gas physics is due, in part, 
to the possibility of measuring atomic density profiles by fluorescence 
or absorption imaging. Adopting the language of second quantization,
the corresponding observable is the intensity or density operator,
\begin{equation}
    \hat n( {\bf x} ) = \hat \Psi^\dag( {\bf x} )\hat \Psi( {\bf x} ),
    \label{eq:def-density}
\end{equation}
integrated along a line of sight.  (We shall omit this integral in the
following.)  An `image' thus corresponds to a set of measured
densities $\left\{ \rho( {\bf x} ) | {\bf x} \right\}$.  Averaging
over many images, one gets an estimate for the expectation value
$\bar \rho( {\bf x} ) = \langle \hat n( {\bf x} ) \rangle$.  If the
detector integrates the density signal over some small, but finite
area $\mathcal{A}_{\rm px}( s )$ (the `pixel' no.  $s = 1, 2, \ldots M$), one
deals with a discrete set of operators like
\begin{equation}
    \hat n_{s} = \int\limits_{\mathcal{A}_{\rm px}( s )}\!{\rm d}{\bf x} \, \hat n( {\bf x} ).
    \label{eq:def-pixelized-density}
\end{equation}
Correlations between the atomic density in different points are
related to the field operator in the following way, 
using the bosonic commutation relations
\begin{eqnarray}
    \langle \hat n( {\bf x} ) \hat n( {\bf x}^{\prime} ) \rangle & = &
    \langle \hat \Psi^\dag( {\bf x} ) \hat \Psi^\dag( {\bf x}^{\prime} )
    \hat \Psi( {\bf x}^{\prime} ) \hat \Psi( {\bf x} ) \rangle
    \nonumber\\
    && {}
    + \langle \hat n( {\bf x} )  \rangle
    \delta( {\bf x} - {\bf x}^{\prime} ).
    \label{eq:ordered-density-correlations}
\end{eqnarray}
This quantity defines the density correlation function
\begin{equation}
    \mathcal{P}( {\bf x}, {\bf x}' )
    = \langle \hat n( {\bf x} ) \hat n( {\bf x}^{\prime} ) \rangle -
    \langle \hat n( {\bf x} ) \rangle \langle \hat n( {\bf x}^{\prime}) 
    \rangle
    \label{eq:def-density-corrs}
\end{equation}
that will play a key role in this paper.

Let us illustrate these concepts for the case that the quantized field
operator $\hat \Psi( {\bf x} )$ can be reduced to a single mode.  This
is a common approximation at low temperatures where a macroscopic
fraction of atoms condenses into a single spatial wave function.  The
operator $\hat \Psi( {\bf x} )$ is then replaced by a single annihilation
operator $\hat a_{0}$ multiplying a classical complex field $\Phi( {\bf x})$, 
and the average density is given by
\begin{equation}
    \langle \hat n( {\bf x} ) \rangle = N_{0} |\Phi( {\bf x} )|^2,
    \label{eq:ave-density-Poisson}
\end{equation}
where $N_{0} = \langle \hat a_{0}^\dag \hat a_{0} \rangle$ is the number of
atoms in the `condensate mode' $\Phi( {\bf x} )$ (itself normalized to 
unity).
The density correlations, 
essentially the structure factor of the system~\cite{Pitaevskii2003}, 
are found as
\begin{eqnarray}
    \mathcal{P}( {\bf x}, {\bf x}' )
    &=&
    \left( \Delta N_{0}^2 - N_{0} \right)
	|\Phi( {\bf x} )|^2 |\Phi( {\bf x}^{\prime} )|^2
    \nonumber
\\
    && {}
    + N_{0} |\Phi( {\bf x} )|^2
    \delta( {\bf x} - {\bf x}^{\prime}).
    \label{eq:coh-state-density-corrs}
\end{eqnarray}
The first term, proportional to the Mandel parameter~\cite{MandelWolf}, vanishes
if the system is in an eigenstate of the operator 
$\hat a_{0}$, i.e.,  a coherent state (Poisson statistics). 
The second term describes local fluctuations at the same position 
(for an image: in the same pixel), with a variance that is equal to
the mean density (the mean atom number on the pixel).

It is intuitively clear that when more than a single spatial mode are taken into
account in the field operator expansion, valuable information about
the field's quantum state is hidden in the correlations of the atomic
density \cite{Altman2004}. 
This has
been discussed recently for the Mott insulator--superfluid transition
\cite{Mekhov2007,Mekhov2007a}, and exploited in measurements
on the strongly correlated Mott insulator
phase with ultracold $^{87}$Rb atoms released from an optical
lattice \cite{Folling2005}.
Excited modes also introduce additional fluctuations into an image, 
however. The competition between these two effects will be the 
central theme of this paper. 

\subsection{Parameter estimation}
\label{s:CRB}

\subsubsection{Information measures}

The full joint probability distribution of the atom numbers in every pixel provides 
the complete `counting statistics' of an image. The information content of an image 
is given by the counting statistics via the classical information entropy $I$, which for a 
pixelized image is given by (e.g., see \cite{Refregier2004})
\begin{equation}
    I = -\sum_{\rho_{1}, \ldots \rho_{\Mpx{}}} p( \{ \rho_{s} \} )
    \log p( \{ \rho_{s} \} ).
    \label{eq:general-information}
\end{equation}
In (\ref{eq:general-information}), $p( \{ \rho_{s} \} )$ is the joint probability measure
for the occurrence of detection events with $\rho_{s}$ atoms detected in
pixel ${\rm px}( s )$ ($\Mpx{}$ is the total number of pixels).

A particular application of this information concept appears when
we want to estimate a parameter $q$ ``hidden'' in the image, i.e., 
the counting statistics is
a function $p( \{ \rho_{s} \}; q  )$ of the parameter $q$, referred to as the likelihood 
function (LF) in statistical estimation theory. The hidden parameter can
be the fraction of particles in the condensate mode, the width of
the cloud, or the position of a ``solitonic'' excitation, i.e., the location of a minimum in the
density profile. This will be our example throughout the analysis.
Information processing theory provides an
explicit formula for the optimal signal-to-noise ratio in measuring $q$. 
It can be
translated into a lower limit on the variance, the Cram\'er-Rao bound (CRB),
${\rm Var}( q ) \ge 1/F( q )$. Here, the Fisher information $F( q )$ 
is given by~\cite{Cramer1946,Refregier2004}
\begin{eqnarray}
\label{eq:FisherInfoGen}
F(q) = - \sum_{\rho_{1}, \ldots \rho_{\Mpx{}}} p( \{ \rho_{s} \}; q )
\frac{\partial^2 \log p( \{ \rho_{s} \}; q )
    }{ \partial q^2}.
\end{eqnarray}
In a quantum mechanical framework, the FI  
has a geometrical interpretation as distance (metric), 
depending on a parameter $q$, in the space of 
density operators~\cite{Braunstein1994}.
The multiple sum in Eq.(\ref{eq:FisherInfoGen}) is difficult to
evaluate in general, and we shall focus in this paper on two schemes
where the calculations are feasible: $(i)$ the single-mode approximation for
the field operator, assuming Poissonian atom number counting statistics; 
$(ii)$ a Gaussian approximation for the
probability measure, where the results can be expressed in terms of
the average density and the density correlations. The latter scheme
will be applied to a multi-mode field theory within the Bogoliubov
approximation.

To estimate the parameter $q$ we require a definite prescription of
how to extract it from the data that fluctuate from shot to shot.  
We shall in particular identify an optimal prescription that permits 
to saturate the Cram\'er-Rao bound.

\subsubsection{Poissonian counting statistics}
\label{s:Poisson-statistics}

The single-mode approximation is well known as the mean-field
theory for BEC. We assume here that the variance of condensate
particles is normal, $\Delta N_{0}^2 = N_{0}$ (Poisson statistics)
so that the density correlations are given by the last term in 
Eq.(\ref{eq:coh-state-density-corrs}). The mean field theory is equivalent to a 
Hartree product state Ansatz for the many-body wave function. 
As a result, the probability measure $p( \{ \rho_{s} \} )$
factorizes into Poissonian statistics for each pixel. The summation in
Eq.(\ref{eq:FisherInfoGen}) can then be performed analytically. When
we take the limit of infinitely small pixels, one finds an integral over
the spatial coordinate of the image \cite{Treps2005}
\begin{equation}
    \label{Eqn:FisherPoisson}
    F( q )
    = 4 \int\!{\rm d}x
    \left[\frac{\partial |\Phi(x; q)|}{\partial q}\right]^2,
\end{equation}
where we have made the dependence of the average density
profile $|\Phi(x; q)|^2$ on the parameter $q$ explicit. (From here on,
the complex field $\Phi$ is not normalized to unit norm, but its 
square gives the mean density of the condensate particles.)

The meaning of this formula can be illustrated by a discussion of the
optimal signal processing strategy. Given the image data
$\rho( x; q )$, we construct, as in Refs.\cite{Treps2005,Delaubert2008},
a linear filter $g(x)$ to provide an estimate for $q$:
\begin{eqnarray}
    S(q) &=& \int\!{\rm d}x\, g(x)\,\rho(x; q) \nonumber\\
&\approx&
q\,\int\!{\rm d}x\, g(x)\,\partial_{q }\rho(x; 0),
\label{Eqn:SqGeneral}
\end{eqnarray}
where $g(x)$ is a local gain function on the pixel at position $x$ that
can take positive and negative values.  In the second step we have
assumed, without loss of generality, that the signal vanishes when the
parameter $q$ is zero, and we have performed a Taylor expansion of the
density profile, $\rho(x; q) \simeq \rho(x; 0) +
q\,\partial_q\rho(x; 0)$. The expectation value of the signal, $\bar{S}( q )$, is 
simply obtained in terms of $\bar{\rho}( x; q ) = 
\langle \hat{n}( x ) \rangle$ and its derivative. 

The variance of the signal is found by squaring
Eq.(\ref{Eqn:SqGeneral}) (first line) and expressing the average in
terms of the density correlation function~(\ref{eq:def-density-corrs})
\begin{eqnarray}
\label{Eqn:VarSq}
\Delta S^2&=&\int\!{\rm d}x {\rm d}y \, g(x) g(y)
\mathcal{P}(x,y;q).
\end{eqnarray}
Recalling that different pixels are uncorrelated
[Eq.(\ref{eq:coh-state-density-corrs}) reduces to its last term],
this variance reduces to
\begin{equation}
    \Delta S^2=\int\!{\rm d}x \,g^2(x)\,\bar\rho(x; 0).
    \label{eq:result-signal-variance-uncorr}
\end{equation}
We can now choose the gain function $g( x )$ such that the
signal-to-noise ratio ${\rm SNR} = \bar{S}^2(q)/\Delta S^2$ is maximized.  
This optimization problem has the following solution, as pointed out in
Ref.\cite{Delaubert2008} for a coherent state of light populating a
single spatial mode:
\begin{eqnarray}
\label{Eqn:gOptMF}
g_{\rm opt}(x) = \frac{\alpha}{|\Phi(x; 0)|}
\left(\frac{\partial |\Phi(x; q)|}{\partial q}\right)_{q = 0}.
\end{eqnarray}
Here $\alpha$ is a normalization constant.  This gain
function can also be interpreted as an optimized spatial mode (the
``noise mode'' in the language of Ref.\cite{Delaubert2008}), onto which
Eq.(\ref{Eqn:SqGeneral}) projects the image.
The minimum uncertainty $\Delta q$ for the parameter estimation
corresponds to an SNR of unity, and this reaches the Cram\'er-Rao bound
$(\Delta q)^2 \equiv {\rm Var}(q) = 1/F(q)$ with $F(q)$ given by
Eq.(\ref{Eqn:FisherPoisson}). If the wave function 
$\Phi$ were proportional to $\sqrt{ N_{0} }$, we would find 
a shot-noise limited resolution, $\Delta q \propto 1 / N^{1/2}_{0}$. Due to 
atom-atom interactions, this limit can be overcome, as we shall see.

\subsubsection{Gaussian images}

Given the mean atom
number per pixel $\bar\rho_{s}( q )$ and the (pixelized) density
correlation matrix ${P}_{sj}( q )$ with
[cf.~Eq.(\ref{eq:def-density-corrs})]
\begin{eqnarray}
    \label{Eqn:rhokbar}
    \bar{\rho}_s( q ) &=& \int\limits_{\mathcal{A}_{\rm px}(s)}{\rm 
    d}x\,\rho(x ; q),
\\
    {P}_{sj}( q ) &=& \int\limits_{\mathcal{A}_{\rm px}(s)}\!{\rm d}x
    \int\limits_{\mathcal{A}_{\rm px}(j)}\!{\rm d}y \,
    \mathcal{P}( x, y ; q ),
    \label{eq:def-pixelized-corr-function}
\end{eqnarray}
we may make the assumption of a joint Gaussian probability distribution. Since a Gaussian is fully
characterized by its first and second moments the probability measure is simply given by \cite{Refregier2004}
\begin{eqnarray}
\label{Eqn:GaussianAnsatz}
p({\bfrho} ; q) = \frac{(2\,\pi)^{-\Mpx{}/2}}{\sqrt{{\rm
det}({\mathbf P})}}\,\exp\left[ - \frac{1}{2}\, ({\bfrho} -
\bar{{\bfrho}}) \cdot {\mathbf P}^{-1}\ ({\bfrho} -
\bar{{\bfrho}})\right],\nonumber\\
\end{eqnarray}
where the vector ${\bfrho}$ of length $\Mpx{}$ collects the
detected atom number
variables, $\bar{\bfrho}$ collects the corresponding mean values,
and ${\mathbf P}$ is the covariance matrix with elements ${P}_{sj}( q )$.
The likelihood function $p({\bfrho} ; q)$ depends on $q$ through 
$\bar{\rho}_{s}(q)$ and ${P}_{sj}( q )$.

With the Ansatz~(\ref{Eqn:GaussianAnsatz}) for the LF and
by replacing the sum over discrete particle counts by continuous
integrals, the FI is given by the following analytical
expression \cite{Negretti2008a}
\begin{eqnarray}
\label{Eqn:generalF}
F( q ) &=& \frac{1}{2}\,\left\{\frac{\partial^2_q{\rm det}({\mathbf
P})}{{\rm det}({\mathbf P})}-\left[ \frac{\partial_q{\rm det}({\mathbf
P})}{{\rm det}({\mathbf P})}\right]^2 \right.
\\
&& {} + \left.\sum_{s,j}\left[\frac{\partial^2
(\mathbf{P}^{-1})_{s\,j}}{\partial q^2}\,P_{s\,j} +
2\,(\mathbf{P}^{-1})_{s\,j}\,\frac{\partial\bar{\rho}_s}{\partial q}\,
\frac{\partial\bar{\rho}_j}{\partial q}\right]\right\}.
\nonumber
\end{eqnarray}
If no correlations exist between neighboring pixels,
this expression reduces to Eq.(\ref{Eqn:FisherPoisson}) for the
Poisson case, provided one takes both the limit of small pixel size
and large average atom number per pixel. This is as expected since
for a large average, Poisson and Gaussian statistics become similar.
We discuss an example taken from Ref.\cite{Negretti2008a} below.

Correlations between pixels may increase or decrease the FI and make 
the parameter estimate more or less precise. A detailed calculation is 
discussed in Sec.\ref{s:crb-with-Bogoliubov}.

\subsection{Examples: kinks and vortices}
\label{s:soliton-with-Poisson}

We evaluate now the Fisher information for measurements of the
position of a solitonic excitation in a Bose-condensed gas.
We consider first a 1D setting where the quantum field theory is given
by the following nonlinear evolution equation for the field operator
[the Gross-Pitaevskii equation (GPE)]
\cite{Pethick2002,Pitaevskii2003}
\begin{eqnarray}
\label{Eqn:GPEgen}
i\,\hbar\,\frac{\partial}{\partial t}\hat\Psi =
\left[-\frac{\hbar^2}{2\,m}
\frac{ \partial^2 }{ \partial x^2 }
+ V_{\rm ext} + g\,\hat\Psi^\dag \hat\Psi \right]\,\hat\Psi,
\end{eqnarray}
where $m$ is the atom mass, $V_{\rm ext}$ is an external potential,
and $g$ the binary interaction strength proportional
to the $s$-wave scattering length $a_{s}$.
Within mean-field theory, the field operator is replaced by the
complex order parameter $\Phi(x)$ (the condensate wave function or order parameter)
that also satisfies Eq.(\ref{Eqn:GPEgen}). The mean atom density is then
$\bar\rho(x) = |\Phi(x)|^2$, thus normalizing the order parameter 
to the particle density.

A family of ``dark" solitonic excitations of the Bose condensate
exists for repulsive interactions ($g > 0$).  In the homogeneous case
($V_{\rm ext} = 0$), it is given by \cite{Pitaevskii2003}
\begin{eqnarray}
\label{Eqn:Soliton}
\Phi(x; q)
= \frac{\sqrt{n}}{c}\left\{i\upsilon + \sqrt{c^2 - \upsilon^2}
\tanh\left[\kappa(\upsilon)(x - q)\right]\right\},
\end{eqnarray}
which shows a local minimum in the density at the position $q\equiv q(t) =
q(0) + \upsilon t$. The width of this dip is specified by
$\kappa(\upsilon) = (\sqrt{2}\,\xi)^{-1} \sqrt{1 - \upsilon^2/c^2}$
where $\xi = \hbar/\sqrt{2\,m\,g\,n}$ is the so-called healing length
and $n$ is the one-dimensional ``background'' density. The soliton moves
with constant velocity $\upsilon$ that does not exceed the sound velocity
$c = \sqrt{n\,g/m}$. For a soliton at rest, the density is strictly zero
at $x = q$, and the phase of the condensate wave function increases
by $\pi$ when crossing this point (a ``kink soliton''). The kink 
becomes wider and the density dip disappears as the velocity 
$\upsilon \to \pm c$. 
The wavefunction in (\ref{Eqn:Soliton}) is normalised such that the
atom density asymptotically approaches the constant value of $n$
(actually, after few healing lengths).  Such solitons have been
created in BEC experiments by shining a laser field on one half of the
atomic ultracold cloud, which induced a relative phase by the AC Stark
effect \cite{Burger1999}.

Topological excitations in two dimensions are vortices where the phase
of the wave function increases by a multiple of $2\pi$ when circling
around a zero in the density.  We discuss an example below.

\subsubsection{Dark soliton}

\begin{figure}[t]
\begin{center}
\includegraphics[width=4.9443cm,height=8.0cm,angle=90]{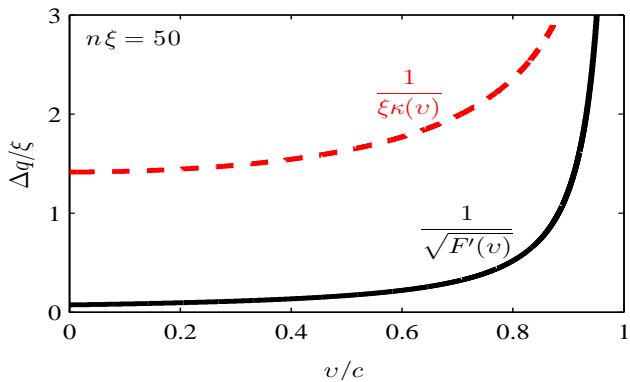}
\end{center}
\caption[]{(Colour online) Soliton position uncertainty $\Delta q$, in units of the healing length 
$\xi$, versus soliton velocity $\upsilon$ ($c$ is the speed of sound). The scaled Fisher information 
is given by $F^{\prime} = F\,\xi^2$.}
\label{fig:Fig1}
\end{figure}

Due to the shot noise fluctuations in the detected atoms the density dip
is washed out and one has to find a good estimate for the kink position $q$.
The formula~(\ref{Eqn:FisherPoisson}) for the FI within Poisson statistics
can be evaluated exactly for the dark soliton given by Eq.(\ref{Eqn:Soliton}).
The result is [recall that Eq.(\ref{Eqn:FisherPoisson}) implies the limit of
infinitely small pixels]
\begin{align}
\label{Eqn:FisherPois}
F &=
 \frac{4\,n \sqrt{2}}{\xi}\,\frac{\upsilon}{c}\,\left\{
 \arctan\left[\frac{\upsilon/c - c/(2\,\upsilon)}{\sqrt{2}
 \,\xi\,\kappa(\upsilon)}\right]
 \right.\\
 &\phantom{=}
 \left. - \arctan\left[\frac{\upsilon/c}{\sqrt{2}\,\xi\,\kappa(\upsilon)}\right]\right\}
 +\frac{8\,n}{3}\left(2 + \frac{\upsilon^2}{c^2}\right)\kappa(\upsilon).\notag
\end{align}
As expected from translation invariance, after integration over
$x$, this does not depend on the soliton position $q$ .  
In Fig.\ref{fig:Fig1} we plot
the minimum uncertainty $\Delta q = F(q)^{-1/2}$ of the soliton position
in units of the condensate healing length, as a function of the
soliton velocity $\upsilon$.
This behaviour makes good sense: when the soliton is almost at rest,
the dip is very sharp and a precise knowledge
of the soliton position can be gathered; when
the velocity approaches $c$, the
dip in the density becomes very shallow, and less information is
available from density measurements.  For comparison, the dashed
curve shows the width $1/\kappa(\upsilon)$ of the density dip: we see
that the measurement precision can largely exceed this value.
The maximum information is
\begin{equation}
    \upsilon = 0: \qquad F =
     \frac{16}{3\sqrt{2}}\frac{ n}{ \xi }
     = \frac{16}{3}\frac{\sqrt{mg}}{\hbar}n^{3/2},
    \label{eq:max-kink-Fisher}
\end{equation}
which implies that, for a soliton at rest, the uncertainty in $q$ scales like $n^{-3/4}$ with
the background density $n$. As pointed out in Ref.\cite{Negretti2008a}, this is a
better scaling than the usual shot noise limit ($\sim n^{-1/2}$).
We emphasize that this enhancement does not require any
squeezed or otherwise entangled multi-atom state. It simply follows
from the shorter wavelengths that occur in a BEC matter wave 
interferometer due to the atom-atom interactions.

The scaling with the density can be understood with the following
statistical argument: the atom number on a pixel of area $\Delta x$ at
position $x$ is given by $N(x) = \rho(x)\,\Delta x \pm
\sqrt{\rho(x)\,\Delta x}$, assuming fluctuations at the shot-noise
level. From error propagation theory the
uncertainty $\Delta q$ on the soliton position $q$ is given by
\begin{equation}
    \Delta q = \frac{ \Delta N(x) }{ |{\rm d}\rho(x)/{\rm d}x| \Delta
    x } \sim \frac{ \xi }{ \sqrt{n\,\Delta x} },
    \label{eq:error-propagation-delta-q}
\end{equation}
where the healing length $\xi$ sets the scale for the density
variation around the dip. By processing different data points across 
the relevant region, this uncertainty can be reduced by a factor
$1/\sqrt{ \Mpx{} }$ where $\Mpx{} \approx \xi/\Delta x$
is the number of pixels across the dip,
which leads to a precision of the order of
\begin{equation}
    \Delta q \sim \sqrt{ \frac{ \xi }{ n  } }
    \propto n^{-3/4}.
    \label{eq:final-precision-1D}
\end{equation}

Let us now turn to the Gaussian approximation to the likelihood
function of an image and discuss the Fisher
information~(\ref{Eqn:generalF}) for the dark soliton. 
We recover the variance and
mean of the Poisson distribution by choosing a covariance matrix with
$P_{s\,j} = {\rm Var}(\rho_s)\,\delta_{s\,j} =
\bar{\rho}_s\,\delta_{s\,j}$, with the variance of the number of atoms
on the $s$-th pixel equal to its mean value $\bar{\rho}_s( q )$.

The integral in Eq.(\ref{Eqn:rhokbar}) with the order parameter given
in (\ref{Eqn:Soliton}) can be worked out analytically and the
derivative with respect to $q$ performed.  The FI (\ref{Eqn:generalF}), 
scaled to $F^{\prime} = F\,\xi^2$, then becomes
\begin{eqnarray}
\label{Eqn:FisherGauss2}
F^{\prime} =
n^2 \xi^2
(1-\upsilon^2/c^2)^2
\left[\sum_s\frac{g_s}{\bar{\rho}_s(0)}
+\frac{1}{2}\sum_s\frac{g_s}{\bar{\rho}_s^2(0)}\right],\nonumber\\
\end{eqnarray}
where we have put $q = 0$ without loss of generality and
\begin{align}
\label{Eqn:g}
\begin{split}
g_s  &= \left\{{\rm sech}^2\left[\kappa(\upsilon)(x_{s} + 
\Delta x)\right]
-{\rm sech}^2\left[\kappa(\upsilon)\,x_s\right]\right\}^2,
\end{split}
\end{align}
\begin{align}
\label{Eqn:rhokbarsol}
\begin{split}
\bar{\rho}_s(0) &= n\,\Delta x - 2\,n\,\xi^2\kappa(\upsilon)
\left\{\tanh\left[\kappa(\upsilon)(x_{s} + 
\Delta x)\right]\right.\\
&\phantom{=}-\left.\tanh\left[\kappa(\upsilon)\,x_s\right]\right\},
\end{split}
\end{align}
with $x_s = s \Delta x$, and $s$ an integer.  
\begin{figure}[t]
\begin{center}
\includegraphics{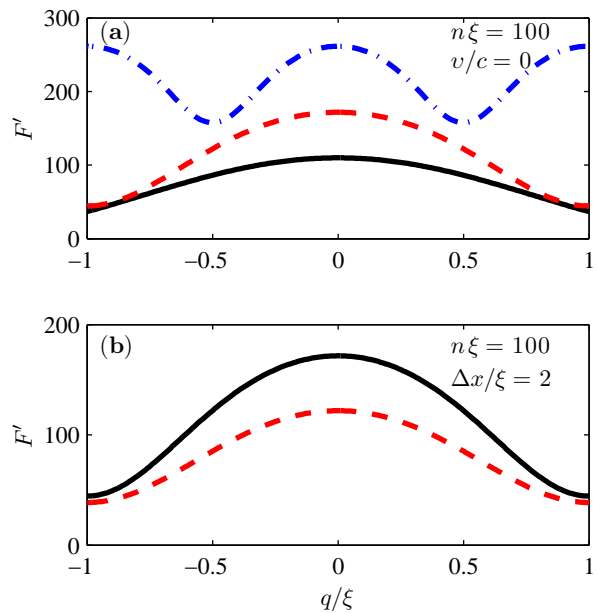}
\end{center}
\caption[]{(Colour online) (a) Rescaled Fisher information as a function of soliton position, $q$, for a
Poissonian LF. The solid (black) line is the result for pixel size $\Delta x/\xi=3$, the dashed (red)
line for $\Delta x/\xi=2$, and the dashdot (blue) for $\Delta x/\xi=1$. (b) Fisher information as a
function of soliton position, $q$, for a Poissonian LF, shown for two different soliton velocities and
fixed $\Delta x$. The solid (black) line is for $\upsilon=0$ and the dashed (red) line is for
$\upsilon=0.3\,c$.}
\label{fig:FisherInfo_qv}
\end{figure}
This expression is giving those pixels a stronger weight where the density profile significantly
changes, as could have been expected. As a consequence, pixels at the outer limits where the atomic
density is flat (already a few healing lengths away from the soliton position $q$) do not contribute
to the FI. We can therefore safely assume that the soliton is located well in the
interior of a detection window for imaging, and even take the limit of an infinitely large window.
(For a discussion of the limit $\Delta x \to 0$, see 
Ref.\cite{Negretti2008a}.)
That is also confirmed in Fig.\ref{fig:FisherInfo_qv}(a): we show the dependence of the FI on the 
soliton position $q$ within a pixel. The oscillations occur at the pixel size $\Delta x$, but for 
$\Delta x\rightarrow 0$, their amplitude becomes smaller and $F^{\prime}$ 
is almost flat.
In Fig.\ref{fig:FisherInfo_qv}(b), we fix the value of $\Delta x$ and
change the soliton velocity $\upsilon$.
Also in that case the FI has a maximum for $q=0$ (soliton at the border between two
pixels).

\subsubsection{Trapped soliton}

The previous discussion shows that the information content is 
concentrated near the soliton position. This suggests that trapped 
solitons, in an inhomogeneous background, behave in a similar way.

Consider first a soliton at rest in a square potential of length $2\ell$.
Assuming a single-mode picture with
Poissonian number statistics, the FI becomes 
\begin{equation}
\label{Eqn:FboxMF}
F^{\prime} = n^2 \,
\xi^2\,
\sum_s\frac{g_s(\Delta x)}{\bar{\rho}_s(0)}\vartheta(\ell - |x_{s}|),
\end{equation}
where the step function $\vartheta(x)$ appears.
As expected, only those pixels covering the confining box contribute
to the information.
In the limit $\ell \to \infty$, we recover the result of an infinite
Bose gas given by the first sum of (\ref{Eqn:FisherGauss2}) with
$\upsilon=0$.  
The same discussion can be applied to the
Gaussian approximation for the complete counting statistics.

In a harmonic trap, the order parameter 
can be approximately written as \cite{Dziarmaga2002}
\begin{eqnarray}
\Phi(x; q)\simeq
\Phi_{\rm bg}(x)\,\tanh\left(\frac{x - q}{\sqrt{2}\,\xi_0}\right),
\end{eqnarray}
where $\Phi_{\rm bg}(x)$ is the Thomas-Fermi (TF) solution for the trap ground state
\cite{Molmer1998}, $\xi_0$ is the healing length for the background
density at the soliton position (it depends on the
density $n_{0} = |\Phi_{\rm bg}(q)|^2$ \cite{Petrov2000}).
Let us focus on solitons at rest, close to the centre of
the trap, $q\approx 0$, and much smaller than the size $R_{\rm TF}$
of the background field. The FI~(\ref{Eqn:FisherPoisson})
then turns out to be
\begin{eqnarray}
\label{Eqn:Fho}
{F}_{\rm trap}( q ) &\simeq&
\frac{ 16 }{ 3 \sqrt{ 2 } }
\frac{ n_{0} }{ \xi_{0} },
\end{eqnarray}
which is very similar to the homogeneous case, keeping the density 
$n_{0}$ the same.

This result can be written in different ways by expressing the central
density $n_{0}$ in terms of other parameters. Taking, for example, the
total number of atoms, we find
\begin{equation}
    F_{\rm trap} \simeq 14.1\,\frac{ N_{0} }{ a_{x}^2 },
    \label{eq:other-Fho}
\end{equation}
where $a_{x}$ is the (single-particle) ground state size in the harmonic
trap. This leads to the usual shot-noise scaling $N_{0}^{-1/2}$ for the
soliton position. Most of the atoms, however, are not needed for the image
since the information content is concentrated near the soliton minimum.
In this respect, the trapped situation is favorable compared to the
homogeneous case since the central density is higher and the healing length 
is smaller in the center than in the condensate wings. This can be made 
quantitative by comparing the harmonic trap to a homogeneous sample of 
the same size and same total number of atoms. The ratio of the Fisher 
Information between the two cases is then
\begin{eqnarray}
\label{Eqn:Fbox}
\frac{ F_{\rm hom} }{ F_{\rm trap} } \simeq 0.11,
\end{eqnarray}
where corrections of order $\xi / R_{\rm TF}$ are neglected, consistent with 
the TF approximation. It should be noted that the size of the system in both 
cases scales in a different way with the atom number.

\subsubsection{Solitons with three-body interactions} 
\label{s:3body-soliton}

The GPE relies on the 
two-body pseudopotential $g\delta({\bf x})$, which describes the interaction 
between the particles. One can also consider low-dimensional Bose superfluids 
in a strong-coupling regime with three-body collisions. We 
analyze here 
the solitonic solution in 
a quasi 1D setup governed by the equation (homogeneous case)
\begin{eqnarray}
\label{Eqn:BeyondGPE}
i\,\hbar\,\frac{\partial}{\partial t}\Phi =
\left[-\frac{\hbar^2}{2\,m}
\frac{ \partial^2 }{ \partial x^2 }
+ \gamma\,|\Phi|^4 \right]\,\Phi
\end{eqnarray}
with the universal coupling constant
$\gamma = (\hbar\pi)^2/(2 m)$. This system has been investigated in detail 
in Refs.\cite{Kolomeisky2000,Brazhnyi2006}. There exists an analytical 
soliton solution for the order parameter
$\Phi(x;q)=\sqrt{n}f(x) e^{i\phi(x)}$, with 
\begin{eqnarray}
f^2(x;q) &=& 1 - \frac{3 [ 1 - (\upsilon / c)^2]}
{2 + \sqrt{1 + 3 (\upsilon / c)^2} \cosh[\kappa^{\prime}(x - q)]}, 
\\
\kappa^{\prime} &=& 2\pi n \sqrt{1 - (\upsilon / c)^2}.
\end{eqnarray}
This kind of solitons
can also appear in a degenerate electron plasma \cite{Shukla2006}, since the equation 
for the order parameter describing the electron density is almost the same. 
As opposed to Eq.(\ref{Eqn:Soliton}), the soliton width $1/\kappa$ is
scaling here $\sim 1/n$ with the density $n$,
and we therefore expect an improvement in the 
soliton kink estimation. Indeed, using the Poisson formula (\ref{Eqn:FisherPoisson}) 
for the Fisher information we obtain for a soliton at rest 
\begin{equation}
    F = 2 \sqrt{3}\pi\ln(2 +\sqrt{3}) \, n^2
    \label{eq:strong-coupling-Fisher}
\end{equation}
that can be compared to Eq.(\ref{eq:max-kink-Fisher}). 
With such topological excitations we thus reach a $1/n$ scaling
as in the Heisenberg limit, without any entanglement or squeezing.

\subsubsection{Vortex line}
\label{s:vortex-imaging}

Finally, we consider a two-dimensional situation with a vortex line. The vortex solution 
of the 3D GPE is described by a core region where the density goes to zero.
Vortices are not stable solutions, however, and only in a frame rotating with a high angular
velocity they correspond to local or global energy minima \cite{Pitaevskii2003}.
Vortices in a rotating BEC with respect to the lab frame
were observed in several experiments, as for example in Ref.\cite{Madison2000}.

We assume the simple case of a Bose gas with uniform confinement along
the $z$ axis (length $L$) and homogeneous in the $xy$-plane. The solution of the
GPE for a gas rotating around the
$z$ axis is given by \cite{Pitaevskii2003}
\begin{eqnarray}
\label{Eqn:VortexGen}
\Phi({\bf r}; q) = \sqrt{n/L}\,f(x,y,z;q)\,
\left( \frac{ x - q + {\rm i} y }{ x - q - {\rm i} y } \right)^{s/2},
\end{eqnarray}
where $n$ is the two-dimensional density, $f$ is a real function,
$(q,0)$ is the position of the vortex line in the $xy$-plane, and $s$
is an integer (the `winding number' or `topological charge').  The order
parameter~(\ref{Eqn:VortexGen}) is an eigenfunction of the angular
momentum with eigenvalue $\hbar\,s$.  Since we are interested in the
calculation of the FI~(\ref{Eqn:FisherPoisson}), which
depends only on the absolute value $|\Phi|$, the angular dependence is
not relevant in the present setting.  In the following we specialise
to the case of a singly charged vortex $s = \pm 1$.  The function $f$
can then be well approximated by \cite{Pethick2002}
\begin{eqnarray}
\label{Eqn:Vortex-f}
f(x,y,z;q) \simeq \frac{\sqrt{(x-q)^2+y^2}}{\sqrt{2\,\xi^2 + (x - q)^2 + y^2 } }.
\end{eqnarray}
We generalize Eq.(\ref{Eqn:FisherPoisson}) for the FI to
this two-dimensional setting and obtain
\begin{align}
\label{Eqn:FisherP-Vortex}
\begin{split}
F &= 4 \int\limits_{0}^{L}\!{\rm d}z \int\!{\rm d}x{\rm d}y
\left[\frac{\partial |\Phi(r,\varphi,z; 0)|}{\partial q}\right]^2 =
\pi\,n,
\end{split}
\end{align}
where the integration is easy to perform in cylindrical coordinates.
It is interesting to note that for a vortex line, the FI
depends on the (two-dimensional) density $n$ linearly. Hence, the fact that 
the vortex core scales with the healing length
does not improve the shot-noise limit for its position detection.
This can be understood from the simple argument of
Eq.(\ref{eq:final-precision-1D}) because the relevant number of atoms
scales with $n \xi^2 \sim \hbar^2 / (2 m g_{2D} )$ which is only
weakly dependent on the density: the dimensionality of the topological
excitation plays a crucial role.

\subsubsection{Dimensionality crossover}
\label{s:crossover}

The last example has shown that the dimension of the topological excitation is very 
important. Coming back to the quasi-1D situation, we discuss here 
briefly the crossover of the nonlinear terms in the GPE as one changes 
the confinement transverse to the long trap axis. 
The 1D solitons  described by Eq.(\ref{Eqn:GPEgen})
are difficult to realize in an experiment because one
needs a high ratio between radial and axial trap frequency. When 
transverse effects are taken 
into account with more accuracy, the GPE is modified into a 1D non-polynomial 
Schr\"odinger equation~\cite{Salasnich2002,Gerbier2004,Theocharis2007,Mateo2008}. 
This changes, for example, the oscillation frequency of a dark soliton 
in a harmonic trap~\cite{Theocharis2007}. In the limits of weak and strong 
coupling, the nonlinearity in the effective GPE becomes polynomial, 
and analytical results for solitonic solutions (bright and dark) can 
be given~\cite{Salasnich2002}.
By inspecting these dark solitons, it is easy to show that the Fisher 
information scales, in the strong coupling regime, as $n^2$ with the 
background density $n$, identical to the scaling found in Sec.\ref{s:3body-soliton}.
This beats the shot noise scaling ($F \propto n$) and illustrates that also 
with the help of two-body interactions, one can reach the scaling of 
the Heisenberg limit in a strongly interacting system.

\section{Quantum field theory for images}
\label{s:qft}

\subsection{Motivation}
\label{s:mft}

Mean field theories are ubiquitous in physics and play an important
role in the explanation of many phenomena in condensed matter physics
\cite{Huang1987,Auerbach1994}, e.g. superconductivity in metals, and
even in high energy physics as for example the quark condensate in the
so called instanton ensemble in QCD (Quantum Chromodynamics)
\cite{Shuryak2004}.  The mean field approach describes successfully
key features of quantum degenerate dilute gases
\cite{Pethick2002,Pitaevskii2003}, for example 
the static, dynamic, and
thermodynamic properties of trapped Bose-Einstein condensates 
\cite{Dalfovo1999}, and it
has been well confirmed in a number of experiments (see, e.g.,
Ref.\cite{Hau1998}).  The mean field approximation applied to an ultracold
Bose gas leads to the Gross-Pitaevski equation~(\ref{Eqn:GPEgen}).
The stationary solutions of the GPE represent the
\emph{macroscopically} occupied spatial mode functions when the
temperature of the trapped gas is well below the critical transition
temperature.  This concept can be formally translated, in the $U(1)$
symmetry broken approach, in the following way: the total matter field
is split into $\hat{\Psi}(x) = \Phi(x) +\delta\hat{\Psi}(x)$, where
$\delta\hat{\Psi}(x)$ represents quantum fluctuations around the
Gross-Pitaevskii solution $\Phi(x)$.  The mean field approach
essentially neglects the fluctuation field.

Mean field methods, however, assess only the average atomic density
and do not provide information about quantum noise correlations, which
are important in the understanding of quantum phase transitions in
ultracold atoms, e.g., anti-ferromagnetic structures or charge density
waves \cite{Bloch2005}.

\subsection{Bogoliubov approximation}
\label{s:Bogoliubov}

\subsubsection{Phonon and zero modes}

Even at zero temperature, the particle density shows fluctuations
that lead to the phenomenon of ``quantum 
depletion''~\cite{Pitaevskii2003,Lifshitz2002,Castin2001}.
This can be described by linearizing
the total many-body problem around the mean field solution.
In other words, one expands the second quantised Hamiltonian
up to quadratic terms in the fluctuation field $\delta\hat{\Psi}$. 
The dynamics of the latter is then
generated by the Bogoliubov-de Gennes linear operator
\begin{eqnarray}
\label{Eqn:Lgp}
\LBdG{} = \left(
\begin{array}{cc}
\HGP{} + g\, |\Phi|^2 & g\, \Phi^2\\
-g\, \Phi^{*2} & -\HGP{} - g\, |\Phi|^2
\end{array}
\right),
\end{eqnarray}
where we have introduced the Gross-Pitaevskii Hamiltonian
\begin{eqnarray}
\label{Eqn:Hgp}
\HGP{} = -\frac{\hbar^2}{2\,m}\frac{\partial^2}{\partial x^2} + V_{\rm ext}(x)
+ g\,|\Phi(x)|^2 - \mu,
\end{eqnarray}
and $\mu$ is the chemical potential. 

The mode expansion of the fluctuation operator $\delta\hat{\Psi}$ is 
complicated by the fact that $\LBdG{}$ has eigenvectors with 
eigenvalue zero. These arise from continuous symmetries that leave 
invariant the energy of the Bose-Einstein order parameter $\Phi$~%
\cite{Huang1987}. A 
well-known example is the $U(1)$ global phase invariance of the 
GPE~(\ref{Eqn:Hgp}), another one is the translation in space of the 
soliton solution~(\ref{Eqn:Soliton}) for a homogeneous system.  
By choosing a dark soliton solution
with a definite phase $\theta$ and a given position $q$, 
one has spontaneously broken these symmetries. 
According to the
non-relativistic Goldstone theorem \cite{Lange1966,Auerbach1994}, 
this spontaneous symmetry breaking is associated with gapless
excitation modes, the Goldstone modes, for any 
system of particles with finite range interactions.
The Goldstone modes produce quantum fluctuations of the
phase and position of the soliton order parameter, even at zero temperature.
The global phase is
conjugate to the number of particles, and the phase
fluctuations can indeed be interpreted as a consequence of the
variations of the chemical potential with particle number
\cite{Molmer1998}.

The eigenvalues $E_{k}$ of the Bogoliubov-de Gennes operator~(\ref{Eqn:Lgp}) 
correspond to eigenvectors $(\ket{u_k},\ket{v_k})$ that we denote 
``phonon modes''. They are normalized and orthogonal according to
\begin{equation}
    \delta_{k p} =
    \int\!{\rm d}x\left\{
    u_{k}^{*}( x ) u_{p}( x ) -
    v_{k}^{*}( x ) v_{p}( x )
    \right\}
    \label{eq:mode-phonons-norm}
\end{equation}
and are accompanied by partner modes $(\ket{v_k}^*, \ket{u_k}^*)$ with 
negative norm and eigenvalue $-E_{k}^*$. This construction fails for 
the so-called zero modes $(\ket{u_{\alpha}}, \ket{v_{\alpha}})$ that 
are in the kernel of $\LBdG{}$~\cite{Lewenstein96a,Castin1998,Castin2001}.
For the global $U(1)$ symmetry, the zero mode is $(\ket{\Phi},-\ket{\Phi^*})$ 
that is generated by applying $- i \partial/\partial\theta$ to 
$(\ket{\Phi}e^{ i\theta },\ket{\Phi^*}e^{ -i\theta })$. The 
norm~(\ref{eq:mode-phonons-norm}) of this mode is zero.
We need partner or adjoint modes $(\ket{u_{\alpha}^{\rm ad}},
\ket{v_{\alpha}^{\rm ad}})$
to saturate the completeness relation in 
the space of fluctuation fields. They can be constructed by
solving the equation $\LBdG{}(\ket{u_{\alpha}^{\rm ad}},
\ket{v_{\alpha}^{\rm ad}}) = m_{\alpha}
(\ket{u_{\alpha}}, \ket{v_{\alpha}})$. The generalized eigenvalue 
$m_{\alpha}$ is called the ``effective mass''. One gets zero modes of 
$\LBdG{}^2$ and a completeness relation in the form~\cite{Castin1998}
\begin{align}
\begin{split}
\label{Eqn:delta}
\delta(x-y) 
&=
\sum_k \{ u_k(x)\,u_k^*(y) - v_k^*(x)\,v_k(y) \}
\\
&\phantom{=}
{} + \sum_{\alpha} \{ 
u_{\alpha}(x)\,[u_{\alpha}^{\rm ad}(y)]^* 
-
v_{\alpha}^{\rm ad}(x)\,v_{\alpha}^*(y)
\},
\end{split}
\end{align}
where $\alpha$ enumerates all broken symmetries and the
adjoint modes are normalized such that
\begin{equation}
    \delta_{\alpha\beta} =
    \int\!{\rm d}x\left\{
    u_{\alpha}^{{\rm ad}*}( x ) u_{\beta}( x ) -
    v_{\alpha}^{{\rm ad}*}( x ) v_{\beta}( x )
    \right\}.
    \label{eq:mode-adjoint-normal}
\end{equation}
In our one-dimensional soliton system, we expect two Goldstone 
modes associated with the global soliton phase $\theta$ and its 
position $q$. The corresponding quantum fluctuations will prove 
important for our analysis. 
We give the phonon and zero modes for this geometry
in Appendix~\ref{a:soliton-excitations}.

The fluctuation operator $\delta\hat{\Psi}(x)$ around the
Gross-Pitaevskii solution $\Phi(x)$ is expanded as
\begin{align}
\begin{split}
\label{eq:deltapsi-expansion}
\delta\hat{\Psi}(x)
& =
\sum_{k} \{
\hat{b}_{k} u_{k}( x ) + \hat{b}_{k}^\dag v_{k}^*( x ) \}
\\
&\phantom{=} {} +
\sum_{\alpha} \{
\hat{P}_{\alpha} u_{\alpha}^{\rm ad}( x )
- i\,\hat{Q}_{\alpha} u_{\alpha}( x )
\}
.
\end{split}
\end{align}
The operators are constructed to implement the commutation relation
$[\delta\hat{\Psi}(x),\delta\hat{\Psi}^{\dagger}(y)] = \delta(x-y)$:
$\hat b_{k}$, $\hat b_{k}^\dag$ are bosonic annihilation and creation operators, 
$[\hat{b}_k,\hat{b}^{\dagger}_p]=\delta_{k\!p}$, and for all broken 
symmetries in the problem, the ``position''
and ``momentum'' operators $\hat Q_{\alpha}$, $\hat P_{\alpha}$ are 
canonically conjugate, 
$[ \hat{Q}_{\alpha}, \hat{P}_{\beta}] = i\delta_{\alpha\!\beta}$. 
The completeness and orthogonality relations~(\ref{Eqn:delta}, 
\ref{eq:mode-adjoint-normal}) are compatible with the construction
\begin{eqnarray}
\hat{Q}_{\alpha} &=& i\,\int\!{\rm d}x \left\{
    u^{{\rm ad}\,*}_{\alpha}( x ) \delta\hat{\Psi}( x ) - v^{{\rm ad}\,*}_{\alpha}( x )
    \delta\hat{\Psi}^\dag( x )
    \right\},
\label{eq:get-zero-operator}
\\
\hat{P}_{\alpha} &=& \int\!{\rm d}x \left\{
    \delta\hat{\Psi}^\dag( x ) u_{\alpha}( x ) -
    \delta\hat{\Psi}( x ) v_{\alpha}( x )
    \right\}.
\label{eq:get-zero-momentum}
\end{eqnarray}
The operator $\hat{P}_{\alpha}$ can be identified with the generator
of the symmetry transformation behind the corresponding zero mode.

\subsubsection{Mean density and correlations}
\label{s:corrdensity}
\label{s:Bogocorrelations}

We now provide a general framework for the mean atomic density and the
density correlations within Bogoliubov theory.  Going beyond
Ref.\cite{Negretti2008a}, we consider the case of phonon modes at
finite temperature.  The field expansion~(\ref{eq:deltapsi-expansion})
is useful for the computation of expectation values because 
different modes are not correlated. 

Evaluating the average density, we get
\begin{eqnarray}
\langle\hat{\Psi}^{\dagger}(x)\hat{\Psi}(x)\rangle &=&
|\Phi(x)|^2 + \sum_{k} \left( 1 + \langle\hat b_k^{\dagger}\hat b_k\rangle\right)|v_{k}(x,q)|^2
\nonumber
\\
&& {}
+\sum_{k} \langle\hat b_k^{\dagger}\hat b_k\rangle |u_{k}(x,q)|^2+ \mathcal{Z}(x),
\label{eq:ave-density-BdG}
\end{eqnarray}
with the occupation number
$\langle\hat b_k^{\dagger}\hat b_k\rangle = 1 / 
(e^{\beta E_k} - 1)$. For thermal states, $\langle \hat 
b_{q} \rangle = 0$.
The so-called quantum depletion is related to 
the finite contribution $\sim |v_{k}|^2$ of phonon modes even at zero
temperature: this is a direct manifestation of quantum density
fluctuations.  Finite temperature adds thermal contributions to the
phonon modes. 

The zero mode contribution is obtained as
\begin{eqnarray}
    && \mathcal{Z}(x) =  \sum_{\alpha} \left\{
|u_{\alpha}^{\rm ad}( x )|^2\,\langle\hat{P}_{\alpha}^2\rangle
+ |u_{\alpha}( x )|^2 \,\langle\hat{Q}_{\alpha}^2\rangle\right.
\label{eq:Dq-gen}
\\
&& \phantom{=}\left.
-{\rm Re}\,[ u_{\alpha}^*( x )\,u_{\alpha}^{\rm ad}( x ) ]
-{\rm Im}\,[ u_{\alpha}^*( x )\,u_{\alpha}^{\rm ad}( x ) ]
\langle \{
\hat{P}_{\alpha}, \,\hat{Q}_{\alpha} \} \rangle\right\},
\nonumber
\end{eqnarray}
where $\{ \cdot, \cdot \}$ denotes the anticommutator.
For the global $U(1)$ symmetry, 
the relevant operator averages are:
$\langle \hat P_{\theta}^2 \rangle = N_{0}$,
$\langle \hat Q_{\theta}^2 \rangle = 1 / (4\,N_{0})$, and
$\langle \{
\hat{P}_{\theta}, \,\hat{Q}_{\theta} \} \rangle = 0$,
they follow from the assumption that 
the condensate mode is in a coherent state with $N_0$ 
particles on average.
The state for the Goldstone mode associated with soliton displacement
is discussed  in Sec.\ref{sec:zero_modes} below. For the moment, we 
only assumed that the operators $\hat Q_{\alpha}$, $\hat P_{\alpha}$ 
average to zero which is plausible since they appear only quadratically
in the Hamiltonian. 

When quantum fluctuations around the mean field are taken into account,
different spatial locations can become correlated beyond the level of
Eq.(\ref{eq:coh-state-density-corrs}) because they probe the same 
delocalized excitation modes.
We find the density correlation function $\mathcal{P}(x,y;q)$
[Eq.(\ref{eq:def-density-corrs})]
by a straightforward expansion of the four-point field correlations 
to second order in the fluctuation operator $\delta\hat{\Psi}$
(this is consistent with the Bogoliubov approximation). 
Thus we get for phonons in a thermal state 
\begin{align}
\begin{split}
    \mathcal{P}(x,y;q) 
    &\simeq
    \left\langle [
    \Phi^*( x ) \delta\Psi( x )
    +
    \delta \Psi^\dag( x ) \Phi( x )
    ] \right.\\
    &\qquad \left. {} \times
    [
    \Phi^*( y ) \delta \Psi( y )
    +
    \delta \Psi^\dag( y ) \Phi( y )
    ]
    \right\rangle\\
    &= \sum_k \left\{
    \left( 1 + \langle\hat b_k^{\dagger}\hat b_k\rangle\right)
    f_{k}( x )f^{\prime}_{k}( y ) \right.\\
    &\phantom{=} \left. {} + 
    \langle\hat b_k^{\dagger}\hat b_k\rangle f_{k}( y )f^{\prime}_{k}( x )
    \right\}
    \\
    &\phantom{=} {} 
    +\sum_{\alpha} \left\{
    \langle \hat P_{\alpha}^2 \rangle \eta_{\alpha}( x ) \eta_{\alpha}^*( y )
    + \langle \hat Q_{\alpha}^2 \rangle
    \varphi_{\alpha}( x ) \varphi_{\alpha}^*( y )
    \right\}
,
\label{eq:result-density-correlations}
\end{split}
\end{align}
where we introduced the functions
\begin{align}
\begin{split}
    f_{k}( x ) = \Phi^*( x ) u_{k}(x) + v_{k}(x) \Phi( x )
    \\
    f^{\prime}_{k}( x ) = \Phi^*( x ) v^*_{k}( x ) + u^*_{k}( x ) \Phi( x )
\end{split}
\end{align}
for the phonons, and the abbreviations
\begin{align}
\begin{split}
    \eta_{\alpha}( x ) &= \Phi^*( x ) u^{\rm ad}_{\alpha}( x ) +
    u^{{\rm ad}*}_{\alpha}( x ) \Phi( x )
    \\
    \varphi_{\alpha}( x ) &= i \Phi^*( x ) u_{\alpha}( x ) - i
    u^{*}_{\alpha}( x )\Phi( x )
    \label{eq:def-zero-mode-fcns}
\end{split}
\end{align}
that appear like ``mode functions'' for the zero mode operators.
We have simplified Eq.(\ref{eq:result-density-correlations}) by 
noting that 
${\rm Im}\,[ u_{\alpha}^*( x )\,u_{\alpha}^{\rm ad}( x ) ] = 0$ 
for the zero mode functions given in 
Appendix~\ref{a:soliton-excitations}.
To proceed, we have to specify the quantum state of the soliton 
displacement modes. 

\subsubsection{Quantum statistics of soliton position}

\label{s:soliton-with-Bogoliubov}
\label{sec:zero_modes}

In the second-quantized many-body theory, the soliton position is
described by an operator $\hat Q_{q}$ whose fluctuations ``fill'' the
density dip, as discussed by
Dziarmaga~\cite{Dziarmaga2002,Dziarmaga2004}. In a homogeneous 
geometry, the dynamics of $\hat Q_{q}$ is similar to a free particle, 
the corresponding Hamiltonian in the Bogoliubov approximation being
$\hat P_{q}^2 / 2 m_{q}$. The stationary states of this Hamiltonian
are momentum eigenstates where the soliton position is maximally 
uncertain. However, the effective mass occurring here is negative, 
$m_{q} = - 4 m ( n / \kappa )$. This mimicks the negative kinetic 
energy of the classical (non-quantized) moving 
soliton~\cite{Muryshev1999,Busch2000b}. 
A similar phenomenon occurs in a harmonic trap where the soliton
displacement mode has a negative frequency (the mode is called
anomalous)~\cite{Dziarmaga2002}. These degrees of freedom are 
therefore
thermodynamically unstable, and we cannot use thermal statistics.
Dziarmaga in Ref.\cite{Dziarmaga2002} has suggested an alternative approach
to specify the quantum state of this degree of freedom, to be used
as initial condition for the subsequent dynamics. The idea is to ``pin''
the soliton to the position $q$ by minimizing the density 
$\hat \Psi^\dag( q ) \hat \Psi( q )$,  which is a quadratic form in 
$\hat P_{q}$ and $\hat Q_{q}$ [see Eq.(\ref{eq:Dq-gen}) at $x = q$].
As shown in Ref.\cite{Negretti2008a},
one finds in this way a Gaussian state
similar to the ground state of a harmonic oscillator with 
respect to the ``Hamiltonian'' (in quotes since it has not the 
dimensions of energy):
\begin{equation}
    \hat h( \hat P_{q}, \hat Q_{q} ) = 
    \frac{ 1 }{ 16 n } \hat P_{q}^2 + n \kappa^2 \hat Q_{q}^2 - 
    \frac{ \kappa }{ 4 },
    \label{eq:zero-mode-Hamiltonian}
\end{equation}
where the mode functions of Appendix~\ref{a:soliton-excitations}, 
Eqs.(\ref{eq:zero-modes}, \ref{eq:adjoint-modes}) have been used. 
The ground state of this 
Hamiltonian gives $\langle \hat h \rangle = 0$ and can also be written 
as the vacuum state corresponding to the annihilation operator
\begin{equation}
    \hat b_{q} = \frac{ - i }{ \sqrt{ 8 n \kappa } }
    \hat P_{q} - 
    \sqrt{ 2 n \kappa } \, \hat Q_{q}.
    \label{eq:q-mode-annihilator}
\end{equation}
With the help of $\hat b_{q}$ and its conjugate operator, we can write
the expansion of $\delta \hat \Psi( x )$ in the same form as for the 
phonon modes. 

\begin{figure}[t]
\begin{center}
\includegraphics[width=4.6353cm,height=7.5cm,angle=90]{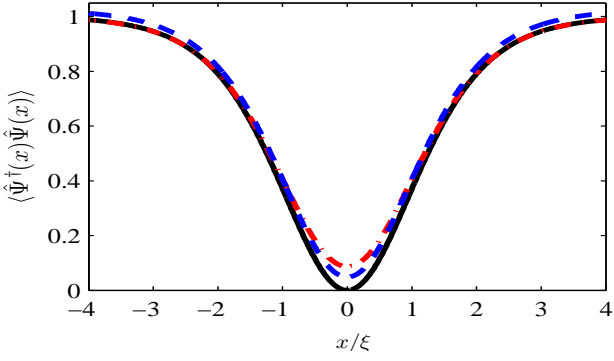}
\end{center}
\caption[]{(Colour online) Density profiles in the notch: the solid (black) line 
corresponds to the squeezing parameter $\zeta=1$ whereas the dashdot (red) to 
$\zeta=100$, while the dashed (blue) line corresponds to the thermal state with 
parameter $\tau = 5/\xi$. Other parameters are: $\ell = 10\,\xi$, $n\,\xi = 100$, 
and $140$ phonon modes have been taken into account.}
\label{fig:FigDen}
\end{figure}

We consider here two classes of states that generalize this ground 
state: ``squeezed'' states and ``thermal'' states. The squeezed state 
depends on the positive parameter $\zeta$:
it is defined as a Gaussian state with quadrature variances
\begin{equation}
    \langle \hat P_{q}^2 \rangle_{\zeta} = \frac{ 2 n \kappa }{ \zeta }
    ,
    \qquad
    \langle \hat Q_{q}^2 \rangle_{\zeta} = \frac{ \zeta  }{ 8 n \kappa },
    \label{eq:squeezed-zero-mode-quadratures}
\end{equation}
where $\zeta = 1$ corresponds to the ground state. For this state, we 
get an average density $\langle \hat h \rangle_{\zeta} = (\kappa/8) 
\left( \zeta^{-1} + \zeta - 2 \right) \ge 0$. The density at $x=q$ 
vanishes for $\zeta = 1$, and Fig.\ref{fig:FigDen} actually shows 
the result of Eq.(\ref{eq:ave-density-BdG}), including contributions 
from quantum noise, for that case. 

The thermal state is defined by analogy to the canonical ensemble as a state that 
maximizes entropy at a given mean value of $\hat h$. This mean value is 
given by
\(
    \langle \hat h \rangle_{\tau} = \frac{1}{2}\kappa /
    ( \exp[ \kappa / (2\tau) ] - 1 )
\)
and it is controlled by a parameter $\tau$ with the dimension density.
The quadratures are
\begin{equation}
    \langle \hat P_{q}^2 \rangle_{\tau} = 2 n \kappa \, 
    \coth[ \kappa / (4\tau) ]
    ,
    \qquad
    \langle \hat Q_{q}^2 \rangle _{\tau} 
    = \frac{ \coth[ \kappa / (4\tau) ]  }{ 8 n \kappa }.
    \label{eq:thermal-zero-mode-quadratures}
\end{equation}
We note that these states give average density profiles that resemble
a partially filled dark soliton, as illustrated in
Fig.\ref{fig:FigDen}. The deviations from the ``optimal case'' are
controlled by the parameters $\zeta$ or $\tau$.  The quantum states
constructed here are not stationary states of the Bogoliubov
Hamiltonian, however, and will evolve in time, as discussed 
in Ref.\cite{Dziarmaga2002}.

\subsection{Discussion of results}
\label{s:crb-with-Bogoliubov}

In Fig.\ref{fig:FigDen} we show density profiles corresponding to different
values of the squeezing parameter $\zeta = 1, \, 100$ and for the thermal parameter 
$\tau = 5/\xi$. The plot confirms that the choice $\zeta = 1$ (or when $\tau\rightarrow 0^+$ 
for a ``thermal'' $q$-state) gives the state with minimum depletion that best resembles 
a condensate with a perfect soliton. For $\tau = 5/\xi$ the minimum in the notch is not zero 
(as for $\zeta\neq 1$), but the density at the edges becomes higher than the density of the 
squeezed quantum state, reflecting the ``thermal'' and the long wave nature of the zero $q$-mode 
state.  

For what concerns the density correlations, it is convenient to bring 
them in a form where the $\delta$-correlated term in 
Eq.(\ref{eq:ordered-density-correlations}) that appears due to 
normal ordering, is subtracted. To this effect, we use the
completeness relation~(\ref{Eqn:delta}) to rewrite the sum over 
$u_k(x) u_k^*(y)$ in the correlation 
function~(\ref{eq:result-density-correlations}) as
\begin{equation}
    \mathcal{P}(x,y) = \Phi( x ) \Phi( y ) \left[ \delta(x - y) +
    \mathcal{J}(x,y) \right],
    \label{eq:def-J-x-and-y}
\end{equation}
where the function $\mathcal{J}(x,y)$ is found as
\begin{widetext}
\begin{align}
\begin{split}
    \mathcal{J}(x,y) 
    &= {\rm Re}\left\{\sum_k \left[2\,v_k(x)\,v_k^*(y) 
         + u_k(x)\,v_k^*(y)
	 + v_k(x)\,u_k^*(y)\right]\right\}
    \\
    &\phantom{=} {} +
    4\left\{
    \langle \hat P_{\theta}^2 \rangle
    u_{\theta}^{\rm ad}( x ) u_{\theta}^{{\rm ad}*}( y )
    +
    \langle \hat Q_q^2 \rangle
        u_{q}( x ) u_{q}^{*}( y )
    \right\}
    - \sum_{\alpha} \{ 
         u_{\alpha}(x)\,[u_{\alpha}^{\rm ad}(y)]^* 
	 +
	 u_{\alpha}^{\rm ad}(x)\,u_{\alpha}^*(y)
    \}
    .
    \label{eq:J}
\end{split}
\end{align}
\end{widetext}
\begin{figure}[pbt]
\vspace{0.1cm}
\begin{center}
\includegraphics[width=7.5cm,height=8.6cm,angle=90]{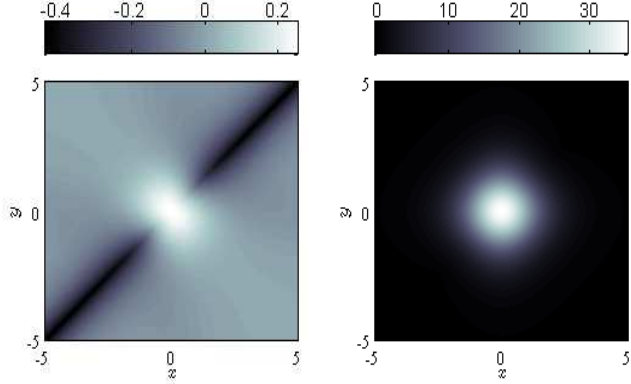}
\end{center}
\vspace{-2.2cm}
\caption[]{Plot of the function $\mathcal{J}(x,y)$ defined in Eq.(\ref{eq:J}).
Parameters: $\ell = 10\,\xi$, $n\,\xi = 100$, and $140$ phonon modes have been taken into account. 
On the left, squeezing parameter $\zeta = 1$ (optimal) and on the right $\zeta = 100$.}
\label{fig:Rfunction}
\end{figure}
In Fig.\ref{fig:Rfunction} we show the behavior of
$\mathcal{J}(x,y)$ for two different values of the squeezing parameter.
Between $\zeta=1$ (left) and $\zeta = 100$ (right), the overall 
magnitude of $\mathcal{J}(x,y)$ changes significantly, demonstrating a 
strong influence of the soliton displacement mode. For a large 
uncertainty in $\hat Q_{q}$ ($\zeta = 100$), the function 
$\mathcal{J}(x,y)$ is positive everywhere and concentrated near
$x \approx y \approx q$. The choice $\zeta = 1$ leads to negative
values (anticorrelations) along the diagonal $x=y$ away from the 
point $x = y = q$. 

\begin{figure}[t]
\begin{center}
\includegraphics[width=6.6cm,height=8.6cm,angle=90]{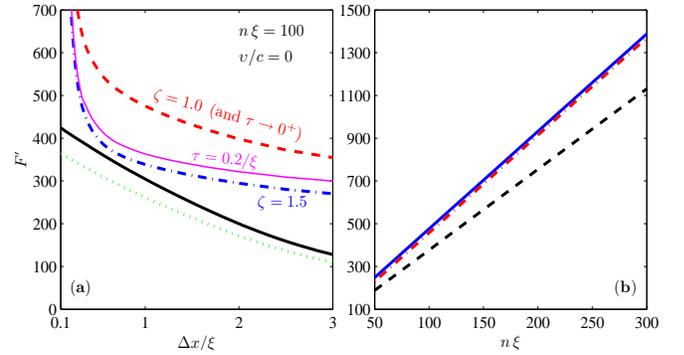}
\end{center}
\vspace{-2.0cm}
\caption[]{(Colour online) 
(a) Rescaled Fisher information $F^{\prime} = F \xi^2$ 
versus pixel size for different regimes: solid (black) line for a Poissonian 
counting statistics in a box and the dotted (green) line in the uniform 
Bose gas limit. 
The other lines take into account quantum and thermal density 
fluctuations, using a Gaussian approximation to the counting 
statistics. They differ in the quantum state of the Goldstone mode
associated with the soliton displacement. 
Dashed (red): state with zero density at soliton position (``squeezing''
parameter $\zeta = 1$, effective ``temperature'' $\tau = 0$);
dashdot (blue): squeezed state $\zeta = 1.5$;
thin solid (magenta): thermal state $\tau = 0.2\,/\xi$.
(b) Rescaled Fisher information versus linear density:
dashed (black) line corresponds to Eq.(\ref{eq:max-kink-Fisher}), and
the dashdot (red) line is the information extracted from the
signal-to-noise ratio for a gain function ${\bf g}^{\sf ao}$ (see
text), and the solid (blue) line is the Fisher information within
Gaussian and Bogoliubov approximations for a finite pixel size 
$\Delta x = 0.7\,\xi$, and $\zeta = 1$. In both pictures the
numerical simulations were made with a box length $2\,\ell = 20\,\xi$,
and by considering 140 phonon modes.}
\label{fig:Comparision}
\end{figure}

The calculation of the Fisher information is done numerically by 
using the phonon and zero modes of
Appendix~\ref{a:soliton-excitations}, normalizing them in a box of
length $2\ell$. 
In Fig.\ref{fig:Comparision}(a), we show the
Fisher information for different approximations: 
a single-mode BEC with Poisson statistics
[dotted line: homogeneous system, 
solid line: in a finite box, Eq.(\ref{Eqn:FboxMF})]
is compared to a multi-mode calculation with a Gaussian counting 
statistics (dashed, dot-dashed, and thin solid lines).
The correlation matrix gives contributions to the FI via the derivatives of
$\bf{P}^{-1}$ and ${\rm det}(\bf{P})$ with respect to the soliton 
position [see Eq.(\ref{Eqn:generalF})]. We have chosen the `optimal dark soliton' 
$\zeta = 1.0$ (dashed line) and a slightly filled (squeezed) one,
$\zeta = 1.5$ (dash-dotted line). The thin solid line corresponds to the 
FI for a thermal state with a parameter $\tau = 0.2/\xi$. 

First, it is interesting to note that one gets more information in a box 
than in the homogeneous case. This is a finite-size effect 
which enhances the impact of ``missing atoms'' in the soliton center.  
Second, the exact result for the Gaussian
multi-mode theory~(\ref{Eqn:generalF})
shows that correlations, including the zero modes, increase the
level of information that we can extract, beyond mean-field theory 
(single-mode approximation). 
Moreover, the dashed curve for
$\zeta = 1$ shows that minimizing the quantum ``filling'' of the notch
provides the highest information. 
The divergence of $F'$ at small pixel size is due to the failure of 
the Gaussian approximation which becomes unphysical, as the average atom 
counts per pixel drop below unity, see Ref.\cite{Negretti2008a}.

In Fig.\ref{fig:Comparision}(b) we show the scaled Fisher information 
versus the scaled density $n$. Both mean-field and multi-mode theory 
give a linear scaling with $n$ which translates into the same power
law $n^{-3/4}$ for the sensitivity of the soliton position. 

In order to understand why the inclusion of noise with the Bogoliubov
description provides an even better resolution than the mean field, we
use the formalism of signal processing introduced in
Sec.\ref{s:Poisson-statistics}.  The signal-to-noise ratio
gives an assessment of the amount of information that
can be extracted from a given statistical estimation strategy.  By
using the completeness relation (\ref{Eqn:delta}) and the result
(\ref{eq:J}) it can be easily shown that the noise 
$\Delta S^2$ [Eq.(\ref{Eqn:VarSq})] splits into
\begin{equation}
    \Delta S^2 = \Delta S_{\rm
    MF}^2 + \Delta S_{\rm ph}^2 - \Delta S_{\rm G}^2
    \label{eq:split-noise}
\end{equation}
for any gain function $g(x)$.  
The three terms here are the contributions of the mean field,
the phonon contribution, and the sum over the 
Goldstone (or zero) modes.  The Goldstone contribution is negative and 
reduces the noise even below the phonon level. 
This is related to the behaviour of the
function $\mathcal{J}(x,y)$ in~(\ref{eq:J}) for 
$\zeta=1$, Fig.\ref{fig:Rfunction}(left): its negative values are
larger in magnitude that its positive ones. Recall that for the 
density correlations~(\ref{eq:def-J-x-and-y}), 
$\mathcal{J}(x,y)$ is multiplied by the
product $\Phi( x ) \Phi( y )$ that is zero for $x = y = q$ and 
increases to a positive constant $n$ for $x=y \ne q$ after a 
few healing lengths. This further enhances the negative contributions 
of $\mathcal{J}(x,y)$, and reduces the noise $\Delta S^2$ relative to 
the mean-field approach.

The above discussion clarifies how the Cram\'er-Rao bound can be smaller
for multi-mode fields compared to the mean field description.
But can one also  identify a strategy to reach the CRB? 
In the Poissonian case, without correlations, the SNR with the optimal 
gain function $g_{\rm opt}( x )$~[Eq.(\ref{Eqn:gOptMF})]
reaches the sensitivity given by the Fisher
information~(\ref{Eqn:FisherPoisson}), and this happens when the 
variance $\Delta S^2\propto \bar{S}(q)$.  
We shall proceed in the same way with the multi-mode theory, but 
work with finite pixel areas to avoid the breakdown of the Gaussian 
approximation. In that case signal and noise are given by 
\begin{align}
\begin{split}
\bar{S}(q) = {\bf g} \cdot 
\frac{\partial\bar\bfrho}{\partial q}
\Delta x,
\\
\Delta S^2 =  {\bf g}\cdot {\bf P}\,{\bf g} \, \Delta x^2,
\end{split}
\end{align}
where the vector ${\bf g}$ (length $\Mpx{}$) represents the
values of the gain function $g( x_{s} )$ on the pixels.  
These expressions are
the discrete versions of Eqs.(\ref{Eqn:SqGeneral}, \ref{Eqn:VarSq}).

We can achieve $\Delta S^2\propto \bar{S}(q)$ by solving the linear system 
${\bf P}\,{\bf g} = \partial_q\bar\bfrho$, giving an optimal gain function
${\bf g}^{\sf ao}$. It is then easy to check that the last term in 
Eq.(\ref{Eqn:generalF}) becomes
\begin{align}
\begin{split}
\sum_{s,j}(\mathbf{P}^{-1})_{sj}\,
\frac{\partial\bar{\rho}_s}{\partial q}\,
\frac{\partial\bar{\rho}_j}{\partial q} =
\frac{\partial\bar\bfrho}{\partial q}
\cdot
{\bf P}^{-1}
\frac{\partial\bar\bfrho}{\partial q}
\propto \bar{S}(q)
.
\end{split}
\end{align}
We were not able to find a similar proportionality 
for the term ${\rm Tr}[(\partial^2{\bf P}^{-1}/\partial
q^2){\bf P}]$. A numerical analysis, illustrated in Fig.\ref{fig:Comparision}(b),
shows that we are nevertheless very close to the CRB 
with the gain function ${\bf g}^{\sf ao}$.

We emphasize that this argument only requires the first and second moments 
of the counting statistics to optimize the SNR. It 
does not directly rely on the Fisher information which is only 
available within the approximation that the entire counting statistics is 
Gaussian. 

\section{Conclusion}
\label{s:conclusion}

We have analyzed in this paper the quantum fluctuations in
the counting statistics of an atomic density image.  We used Bogoliubov
theory, and we observed that zero modes, due to the breaking of the
$U(1)$ and translation symmetries, makes quantitatively a significant
change in the information content of the counting distributions.
We applied our theory to identify the ultimate
information theoretical limits for position measurements of 
dark solitons and other topological excitations in a quasi one-dimensional 
and two-dimensional Bose-Einstein condensate.

In the case of a pure condensate, 
where all particles occupy the same
quantum state described by the GPE, the best
estimation of the soliton position has an uncertainty that scales,
for a weakly interacting system,
as $n^{-3/4}$ with the linear background density $n$. This
beats by a factor $n^{-1/4}$
the scaling of the classical shot noise limit $1/\sqrt{n}$.  We
emphasize that this limit is reached without the need of any
entanglement or squeezing of the system state.  Even more favorable 
scalings are found for strongly interacting systems where the 
nonlinearity appears with a different exponent in the GPE.
In optical and atomic
interferometry, shorter wavelengths provide a better resolution of
phase changes, and we can explain our result as a consequence of the
high wave number content, i.e., the steepness, of the soliton dip.  A
similar improvement could also be obtained by simply applying
fast counter-propagating beams of atoms.  We believe, however, that
the stability properties of solitons and the fact that there is no or
only little net particle current makes this system suitable for
interferometric investigations over longer time scales compared to a 
thermal beam.

We have investigated the influence of quantum fluctuations with
the help of Bogoliubov theory in the weakly interacting case.
The scaling law ($n^{-3/4}$) remains stable, but the prefactor is 
different. The Goldstone modes that are associated to spontaneously 
broken symmetries, in fact increase the sensitivity of the measurements.  
While signal processing theory provides a theoretical limit for the 
sensitivity (the Cram\'er-Rao bound), it generally does not 
provide a method to achieve this limit.  In our approach, 
we were able to find a
nearly optimal filtering function, and showed with a 
signal-to-noise ratio analysis that it almost reaches the 
Cram\'er-Rao bound.


\section*{Acknowledgments}

The authors A.N. and K.M. acknowledge financial support from the
European Union Integrated Project SCALA. K.M. acknowledges 
support of the Multidisciplinary
University Research Initiative ``Quantum Metrology with Atomic 
Systems'', administered by the Office of Naval Research.
C.H. thanks the Deutsche Forschungsgemeinschaft for support (He 2849/3).

\appendix

\section{Excitation modes for a dark soliton}
\label{a:soliton-excitations}

Here we sketch the derivation of the Bogoliubov eigenmodes for a 
system with a soliton in a box of length $2\,\ell$. The analysis 
complements results given in Ref.\cite{Dziarmaga2004}.
We are looking for 
the eigenstates of the Bogoliubov-de Gennes operator~%
(\ref{Eqn:Lgp}) with $V_{\rm ext}(x) = 0$. The background or 
condensate field is given by the wave function 
$\Phi = \Phi(x;q)$ of Eq.(\ref{Eqn:Soliton}). We focus here
on the stationary case  (soliton velocity $\upsilon = 0$).
Simple and compact
expressions for the Bogoliubov eigenmodes can be found in the limit
$\ell \gg \xi$, when the soliton is located well within the
quantization box.  The integral of $|\Phi(x; q)|^2$ over the box
gives the number of condensed atoms $N_{0}$:
\begin{equation}
\label{eq:norm-condensate}
N_{0} 
= 2 \ell n - 2 n / \kappa
\end{equation}
where the negative correction describes the atoms ``missing'' in the 
soliton notch. 
We have set $\kappa \equiv \kappa(\upsilon=0)=1/\sqrt{2}\xi$.
For the sake of simplicity in the notation,
we shall hereafter drop the parametric dependence on $q$ in the order
parameter $\Phi$, the phonon modes $u_k,\,v_k$, and the 
zero modes.

\subsection{Phonon modes}
\label{a:phonon-modes}

Following the approach in Ref.\cite{Busch2000a,Dziarmaga2004}, we find
that the modes of $\LBdG{}$ in a box with periodic boundary conditions
can be written as
\begin{eqnarray}
    \left.
    \begin{array}{c}
	u_{k}( x ) \\
	v_{k}( x )
    \end{array}
    \right\} &=&  
M_{k}
\,e^{i\,k\,x}\,\left\{
\frac{ k }{ \kappa }
{\rm sech}^2[\kappa\,(x-q)] \right.
\label{eq:phonon-modes}
\\
&&  \phantom{=} \left. 
{} +
\beta_{k}^{\pm} 
\left(
\frac{ k }{ 2 \kappa } + i\,\tanh[\kappa\,(x-q)] \right) 
\right\}
,
\nonumber
\end{eqnarray}
where the upper [lower] sign applies to $u_{k}$ [$v_{k}$], respectively.
Here, $M_{k}$ is a normalisation constant [given in 
Eq.(\ref{eq:Mk-result})],
\begin{equation}
    \beta_{k}^{\pm}  = 
\left(\frac{k}{\kappa}\right)^2 \pm
    \frac{2\,E_{k}}{m \, c^2}
    ,
\end{equation}
$c$ is the speed of sound, and the phonon energy is given by
\begin{eqnarray}
    E_{k} & = & \hbar \,c \,|k|\,\sqrt{1 + \frac{k^2}{4\,\kappa^2}}.
\end{eqnarray}
This energy is the same as on a homogeneous background condensate.

The presence of the soliton becomes manifest in the
total phase shift of a phonon passing from the left to the right
end of the box. It is given by $2\,k\,\ell + \Delta\varphi(k)$, where
$\Delta\varphi(k) = 2\,\arctan(2\,\kappa/k)$ is due to
the interaction with the density notch.  
The quantisation condition for the wave number $k$ is thus:
\begin{eqnarray}
\label{Eqn:Quantkn}
2\,k_j\,\ell + \Delta\varphi(k_j) = 2\,\pi\,j \qquad\qquad
j = \pm 1, \pm 2, \ldots
\end{eqnarray}
[The case $j = 0$ is excluded because $|2 k \ell + \Delta\varphi(k)| > \pi$.]
In Sec.\ref{s:Bogoliubov}, sums over $k$ are understood as running over this
discrete set of wavenumbers. 

\begin{figure}[t]
\begin{center}
\includegraphics[width=4.6353cm,height=7.5cm,angle=90]{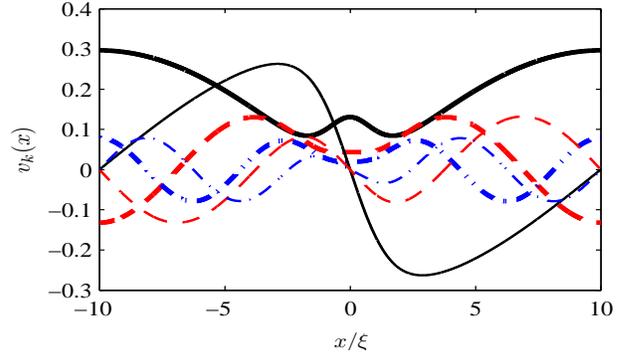}
\end{center}
\caption[]{(Colour online) 
First three Bogoliubov mode functions $v_{k}( x )$ for $q = 0$,
normalized in a box of length $2 \ell = 20\,\xi$ (periodic boundary
conditions).
Thick (thin) lines represent the real (imaginary) part of $v_{k}(x)$, respectively.
Solid (black) line: $k_{1} = 0.5379 \,(\pi /\ell)$,
dashed (red) line: $k_{2} = 1.6093 \,(\pi /\ell)$,
dashdot (blue) line $k_{3} = 2.6704 \,(\pi /\ell)$. These $k$-values
can be compared to a free particle where $k_{j} = 0, \, 1, \, 2 \,(\pi /\ell)$.}
\label{fig:Fig5}
\end{figure}

From the normalization condition~(\ref{eq:mode-phonons-norm}) for the 
phonon modes, 
we obtain the normalization constant $M_{k}$
\begin{align}
\label{eq:Mk-result}
\begin{split}
M_{k} = \frac{\kappa}{2\,k}\,\sqrt{\frac{\kappa\,g\,n}{2\,\epsilon_{k}}}
\,\left\{\ell\,\kappa\,\left[ \frac{k^2}{2\,\kappa^2} + 2 \right]
- 1
\right\}^{-1/2}.
\end{split}
\end{align}
In Fig.\ref{fig:Fig5} we show the first three
eigenfunctions $v_{j}$ with $j=1,\,2,\,3$, for $q = 0$.

\subsection{Zero modes}

For our quantum degenerate Bose gas the Goldstone modes originate from
the breaking of the global $U(1)$ phase symmetry and translational
symmetry by assigning a phase $\theta$ and a value of the displacement $q$ to
the order parameter $\Phi$ respectively.  In order to determine the associated mode
functions,
we simply differentiate the order parameter $\Phi$ with 
respect to its global phase $\theta$ or the parameter $q$, 
as described in Sec.\ref{s:Bogoliubov} and Ref.\cite{Dziarmaga2004}.
We obtain
\begin{eqnarray}
\label{eq:zero-modes}
u_{\theta}( x )
&=&
\Phi( x)
\nonumber\\
u_{q}( x )
&=& - i\,\kappa\,\sqrt{n}\,{\rm sech}^2[\kappa\,(x-q)]
\end{eqnarray}
with $v_{\alpha}( x ) = - u^*_{\alpha}( x )$. 
We have chosen here $\theta = 0$, i.e., a real-valued 
$\Phi( x)$, otherwise the phase of $u_{q}( x )$ would be different. 
With respect to the
Bogoliubov scalar product
Eq.(\ref{eq:mode-phonons-norm}), the modes~(\ref{eq:zero-modes}) have
zero norm and are mutually orthogonal.
The zero modes are accompanied by adjoint vectors that satisfy
$\LBdG{} u^{\rm ad}_{\alpha} \propto u_{\alpha}$ and are found as
\begin{eqnarray}
u_{\theta}^{\rm ad}( x )
&=& \frac{\kappa}{2\,(N_0\,\kappa + n)}\,
\left[
\Phi( x) +
i\, x\, u_{q}( x )
\right],
\nonumber\\
u_{q}^{\rm ad}( x )
&=& \frac{-i}{4\,\sqrt{n}}.
\label{eq:adjoint-modes}
\end{eqnarray}
They have zero norm because
$v_{\alpha}^{\rm ad}( x ) = u^{{\rm ad}*}_{\alpha}( x )$.
The adjoint modes are mutually orthogonal with respect 
to Eq.(\ref{eq:mode-phonons-norm}) which explains the
presence of the second term in Eq.(\ref{eq:adjoint-modes}).
For the normalization and this extra term, we consider the limit 
$\kappa \,\ell \gg 1$ and neglect exponentially small corrections.

\bibliographystyle{apsrev}
\bibliography{NegrettiBEC}

\end{document}